\documentclass[aps,prl,reprint,showpacs,amssymb,amsmath,superscriptaddress]{revtex4-1}
\usepackage[locale=UK,separate-uncertainty=true,uncertainty-separator={\,},multi-part-units=brackets,retain-explicit-plus]{siunitx}
\usepackage{graphicx}
\usepackage{booktabs}
\usepackage{bm}
\usepackage{accents}
\usepackage{braket}
\usepackage[euler]{textgreek}
\usepackage[T1]{fontenc}
\usepackage{lmodern}
\usepackage[utf8]{inputenc}
\usepackage{bibunits}
\usepackage{hyperref}

\DeclareSIUnit\au{{a.u.}}
\AtBeginDocument{
\heavyrulewidth=.08em
\lightrulewidth=.05em
\cmidrulewidth=.03em
\belowrulesep=.65ex
\belowbottomsep=0pt
\aboverulesep=.4ex
\abovetopsep=0pt
\cmidrulesep=\doublerulesep
\cmidrulekern=.5em
\defaultaddspace=.5em
}
\defaultbibliographystyle{apsrev4-1}
\hypersetup{
	pdfpagemode=UseOutlines,
	bookmarksnumbered=true,
	pdflang=en,
	hidelinks,
	hypertexnames=false,
}
\newcommand\fsl[1]{(#1)}
\newcommand{\SupplementPrefix}{S}
\newcommand{\vectLeft}[1]{\accentset{\leftharpoonup}{#1}}
\newcommand{\vectRight}[1]{\accentset{\rightharpoonup}{#1}}
\newcommand\effS{\star}
\newcommand\er{\textrm{\textit{e}}}
\newcommand\eri[1]{\textrm{\textit{e}}_{\textrm{\textit{#1}}}}
\newcommand\egi[1]{\textsf{\textit{e}}_{\textrm{\textit{#1}}}}
\newcommand\Ir{\textrm{\textit{I}}}
\newcommand\Ag{\textsf{\textit{A}}}
\newcommand\IrB{\textrm{\textit{\fontseries{b}\selectfont I}}}
\newcommand\IgB{\textsf{\textit{\fontseries{b}\selectfont I}}}
\newcommand\rt[1]{\textrm{#1}} 
\newcommand\gt[1]{\textrm{#1}} 
\newcommand\gtA[2]{{^{\textrm{#2}}\textrm{#1}}}
\newcommand\rtAI[2]{{^{\textrm{#2}}\textrm{#1}^{\textrm{+}}}} 
\newcommand\rtI[1]{{\textrm{#1}^{\textrm{+}}}}
\newcommand\rtMI[1]{\textrm{#1}_{\textrm{2}}^{\textrm{+}}} 
\newcommand\hyt[1]{\textrm{#1}} 
\newcommand\htA[2]{{^{\textrm{#2}}\textrm{#1}}}
\newcommand\htAI[2]{{^{\textrm{#2}}\textrm{#1}^{\textrm{+}}}}
\newcommand\htI[1]{{\textrm{#1}^{\textrm{+}}}}
\newcommand\eIr{\textrm{\textit{eI}}}
\newcommand\erso{\hspace{1pt}\hat{\hspace{-1pt}\bm{s}}} 
\newcommand{\eIgVi}[2]{\textsf{\textit{#1}}_{\textsf{\textit{#2}}}} 
\newcommand\egso{\hspace{1pt}\hat{\hspace{-1pt}\textsf{\textbf{\textit{s}}}}}
\newcommand\IrAg{{\textrm{\textit{I}}\textsf{\textit{A}}}}
\newcommand\IrAgPo{\hspace{1.5pt}\hat{\hspace{-1.5pt}\bm{P}}} 
\newcommand\erAg{{\textrm{\textit{e}}\textsf{\textit{A}}}}
\newcommand\ArAg{{\textrm{\textit{A}}\textsf{\textit{A}}}}
\newcommand\ArAgNvJ{{N\!v\!J}}
\newcommand\IrAgM{{\textrm{\textit{I}}\textsf{\textit{A}}}}
\newcommand\IrAgS{{\textrm{\textit{I}}\textsf{\textit{A}}}}

\begin{document}
\begin{bibunit}
\title{Rydberg Molecules for Ion-Atom Scattering in the Ultracold Regime}

\author{T. Schmid}
\author{C. Veit}
\author{N. Zuber}
\author{R. L\"ow}
\author{T. Pfau}
\affiliation{5. Physikalisches Institut and Center for Integrated Quantum Science and Technology, Universit\"at Stuttgart, Pfaffenwaldring 57, 70569 Stuttgart, Germany}
\author{M. Tarana}
\affiliation{J. Heyrovsk\'y Institute of Physical Chemistry of the ASCR, v.v.i., Dolej\v{s}kova 2155/3, 182\,23 Prague 8, Czech Republic}
\author{M. Tomza}
\affiliation{Centre of New Technologies, University of Warsaw, Banacha 2c, 02-097 Warsaw, Poland}
\affiliation{Faculty of Physics, University of Warsaw, Pasteura 5, 02-093 Warsaw, Poland}
\date{\today}

\begin{abstract}
We propose a novel experimental method to extend the investigation of ion-atom collisions from the so far studied cold, essentially classical regime to the ultracold, quantum regime.
The key aspect of this method is the use of Rydberg molecules to initialize the ultracold ion-atom scattering event.
We exemplify the proposed method with the lithium ion-atom system, for which we present simulations of how the initial Rydberg molecule wave function, freed by photoionization, evolves in the presence of the ion-atom scattering potential.
We predict bounds for the ion-atom scattering length from \emph{ab initio} calculations of the interaction potential.
We demonstrate that, in the predicted bounds, the scattering length can be experimentally determined from the velocity of the scattered wave packet in the case of $\rtAI{Li}{6}$\,-\,$\gtA{Li}{6}$ and from the molecular ion fraction in the case of $\rtAI{Li}{7}$\,-\,$\gtA{Li}{7}$.
The proposed method to utilize Rydberg molecules for ultracold ion-atom scattering, here particularized for the lithium ion-atom system, is readily applicable to other ion-atom systems as well.
\end{abstract}

\maketitle

\begin{figure}[htbp]
\centering
\vspace{1.5ex}
\includegraphics[width=1.006\columnwidth]{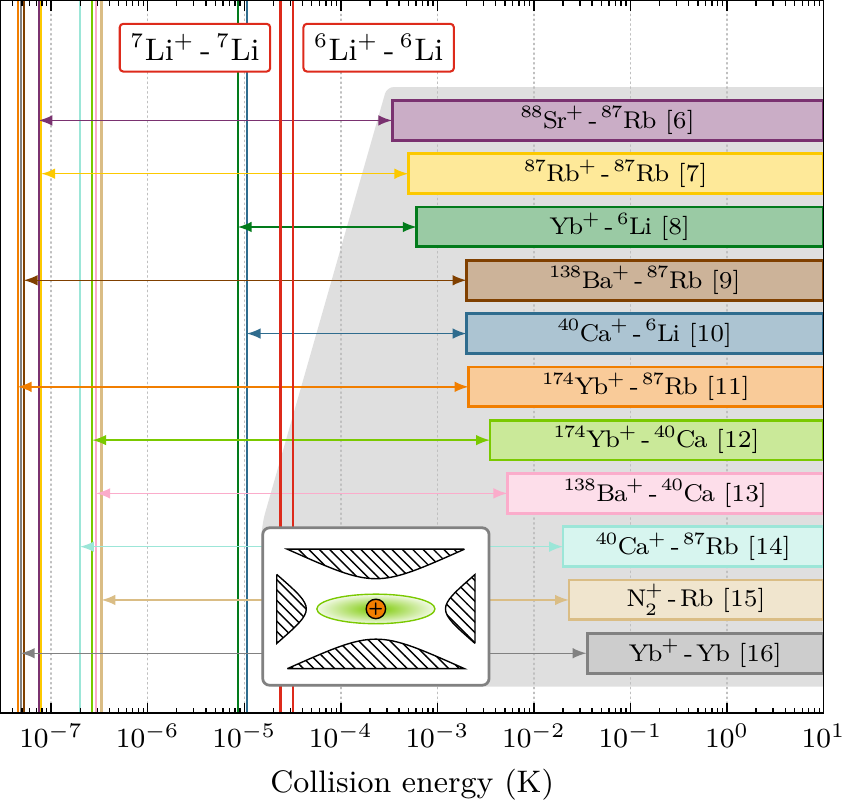}
\caption{
In the past few years, the ion-atom interaction could be studied down to the millikelvin regime for various ion-atom combinations by the use of hybrid traps, with the ion held in a Paul trap (inset).
However, the ultracold, quantum scattering regime could not be reached yet (see color-coded lines for the respective $S$-wave scattering limits $E_0$).
We propose an experimental method to enter the ultracold ion-atom scattering regime using Rydberg molecules.
We demonstrate this method with the lithium ion-atom system which features an early onset of the $S$-wave scattering regime due to its small reduced mass.
}
\nocite{Meir16b,Haerter12,Joger17b,Schmid10,Saito17,Zipkes10a,Rellergert11,Sullivan12,Hall13b,Hall12,Grier09}
\label{fig:Intro}
\end{figure}

The considerable achievements made in the field of degenerate quantum gases over the past decades rely on the exact understanding and control of interactions between neutral atoms in the ultracold regime~\cite{BoseVarenna,FermiVarenna}.
Considering the interactions between ions and atoms, substantial work has been done in the cold, but essentially classical regime, deploying hybrid ion-atom traps.
These hybrid traps combine a Paul trap for the ion with an optical and/or a magnetic trap for the atoms.
Both elastic and inelastic collisions have been studied in these traps for various ion-atom combinations~\cite{Haerter14,Zhang17,Tomza17}.
However, the ultracold, quantum regime, i.e., the $S$-wave collision regime, could not be reached with any of these systems (see Fig.~\ref{fig:Intro}).
Cetina, Grier, and Vuleti\'c showed~\cite{Cetina12} that by the use of a Paul trap there arises a micromotion-induced limit on the minimum collision energy which can be reached.
Only for a combination of a heavy ion with a light atom, e.g., the $\htI{Ca}$\,-\,$\hyt{Li}$ or the $\htI{Yb}$\,-\,$\hyt{Li}$ system, might the $S$-wave collision regime be entered.
For both of these ion-atom combinations, collision measurements have recently been carried out in the millikelvin energy range~\cite{Saito17,Joger17b} which is, however, still in the classical regime.
Schaetz and co-workers~\cite{Huber14a} follow a different path to enter the ultracold regime.
To avoid the spurious heating of the ion by the Paul trap, they optically trap the ion, and they are currently working on the simultaneous optical trapping of the ion and the atoms~\cite{Lambrecht17}.

The generation and characterization of many different types of Rydberg molecules has been an active area of research in the past few years.
These molecules consist of a Rydberg atom and at least one ground state atom which is bound to the Rydberg ionic core at a very large internuclear distance via its attractive interaction with the Rydberg electron~\cite{GreenSadeghpour00}.
Triplet $s$-state Rydberg dimers, trimers, tetramers, and pentamers have been observed~\cite{BendkowskyPfau09,BendkowskyPfau10a,GajPfau14}, and molecular lifetimes have been measured~\cite{Butscher11,Camargo16}.
Furthermore, triplet $d$-state dimers have been studied~\cite{KruppPfau14,AndersonRaithel14} as well as mixed singlet-triplet dimers~\cite{SassmannshausenDeiglmayr15,Boettcher16}.
Finally, trilobite~\cite{LiPfau11,Bellos13,BoothShaffer15,Kleinbach17} and butterfly~\cite{Niederpruem16b} molecules have been investigated.

State-of-the-art \emph{ab initio} calculations could determine the interaction potential of light few-electron systems to such a precision that reliable predictions of the scattering length could be made, e.g., for the scattering of two helium atoms, either in their ground state~\cite{Przybytek10} or in the metastable 2\,$^3S_1$ state~\cite{Przybytek05}.
Accurate estimations of the scattering length could also be given for weakly interacting many-electron systems with a very small reduced mass, e.g., for metastable helium scattering off alkali-metal atoms~\cite{Knoop14,Kedziera15}.
Precisely predicting the scattering length for heavier systems is very challenging, though.

In this Letter, we propose a novel experimental method to extend the investigation of ion-atom interactions from the hitherto studied cold regime to the unexplored ultracold regime.
The key aspect of this method is the unprecedented use of a Rydberg molecule as a tool to initialize an ultracold ion-atom scattering event.
This makes a separate trap for the ion, be it a Paul trap or an optical trap, expendable.
The proposed method allows for the experimental determination of the ion-atom scattering length and thus provides a very valuable benchmark for its \emph{ab initio} calculation.

The starting point of the proposed procedure is the photoassociation of a single Rydberg molecule in an ultracold, dilute atomic cloud.
The Rydberg molecule is then photoionized to start the ultracold ion-atom scattering event between the Rydberg ionic core and the ground state atom; i.e., the initial Rydberg molecule wave function, freed by photoionization, evolves in the presence of the ion-atom scattering potential.
Depending on the scattering length, either the entire scattered wave packet is free and dispersively expanding, or it splits into a free part and a bound part, indicating the formation of a molecular ion.
The detection of the free ion and/or the molecular ion with a time- and position-sensitive single-ion detector concludes the single ultracold ion-atom scattering event.
The frequent repetition of this single scattering event eventually allows for the determination of the ion-atom scattering length.

We exemplify the proposed method to enter the ultracold ion-atom scattering regime with the lithium ion-atom system.
It features an early onset of the $S$-wave scattering regime due to its small reduced mass $\mu$.
Its $S$-wave scattering limit $E_0=(2\mu^2C_4)^{-1}$~\cite{Grier09} ($C_4=\SI{164.2}{\au}$~\cite{Miffre06}) is approximately \num{2} orders of magnitude larger than the respective limits for the ion-atom systems so far studied in the millikelvin range, except for the $\htAI{Ca}{40}$\,-\,$\htA{Li}{6}$ and the $\htI{Yb}$\,-\,$\htA{Li}{6}$ system (see Fig.~\ref{fig:Intro}).
We present an \emph{ab initio} calculation for the interaction potential of the strongly interacting five-electron $\rtI{Li}$\,-\,$\gt{Li}$ system, which is for the first time precise enough to yield usable bounds for the ion-atom scattering length.

\begin{figure*}[htbp]
\centering
\includegraphics[width=\textwidth]{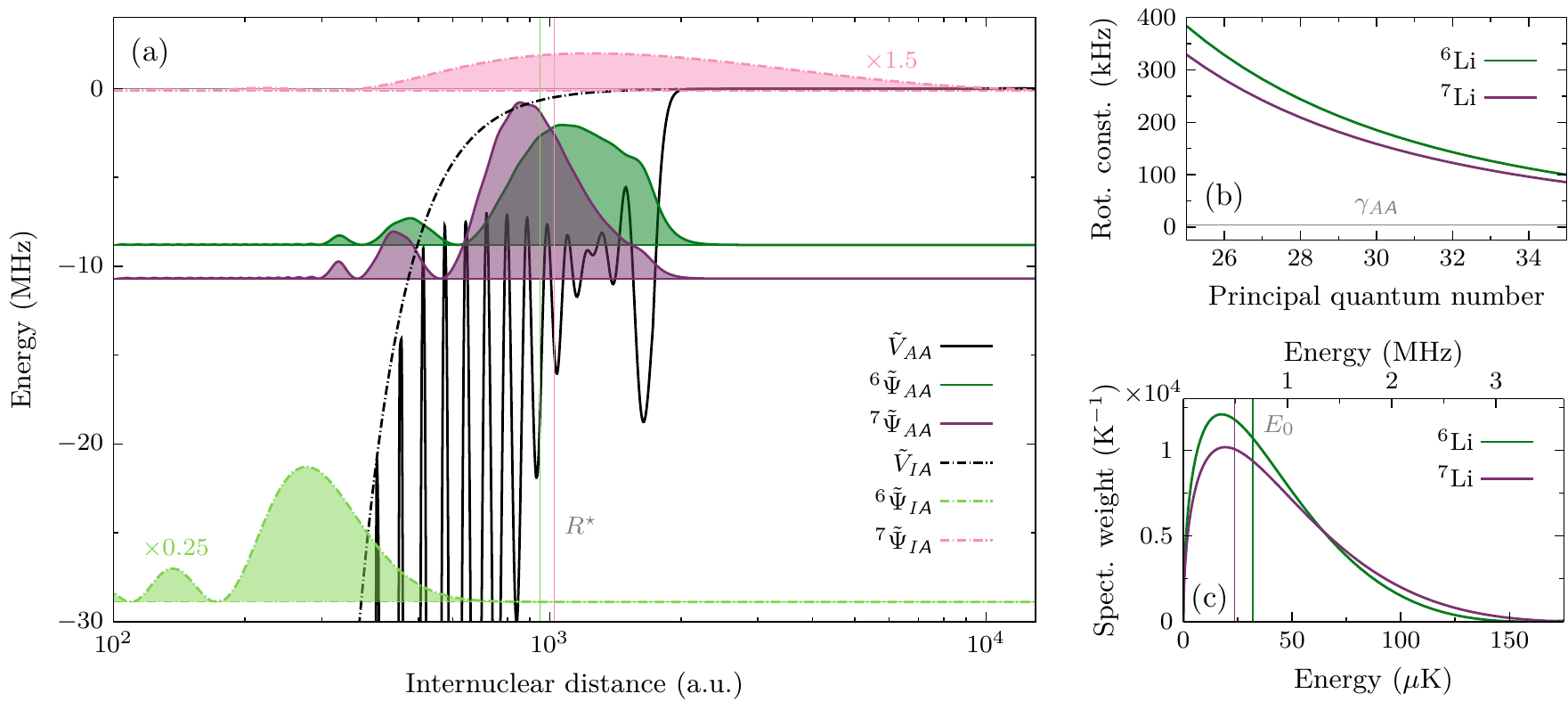}
\caption{
Using Rydberg molecules to initialize an ultracold ion-atom scattering event, exemplified with lithium.
The Rydberg molecule wave function of interest, more precisely, $R^2\left|\tilde{\Psi}_\ArAg(R)\right|^2$, is shown in \fsl{a} for both isotopes.
It is bound in the homonuclear \rt{Li}($30s$)\,-\,\gt{Li}($2s$) potential $\tilde{V}_\ArAg(R)$.
It is in its spherically symmetric rovibrational ground state ($\tilde{v}=\tilde{J}=0$), which can be experimentally addressed, evident from \fsl{b}, as the rotational constant is larger than the Rydberg molecule decay linewidth $\gamma_\ArAg$.
Freed by photoionization, the initial Rydberg molecule wave function evolves in the spherically symmetric ion-atom interaction potential $\tilde{V}_\IrAgM(R)$, with $R^\effS$ denoting its characteristic radius.
Because of angular momentum conservation, only $S$-wave scattering occurs ($\tilde{J}=0$) despite the components above $E_0$ in the Rydberg molecule energy spectrum; see \fsl{c}.
The overlap between $\tilde{\Psi}_\ArAg$ and the last bound molecular ion wave function $\tilde{\Psi}_\IrAgM$ [see \fsl{a}] determines the bound fraction in the scattered wave packet.
}
\label{fig:InitialState}%
\end{figure*}

Rydberg molecules are key to the proposed method in order to initialize the ultracold ion-atom scattering event.
The binding in these molecules is established by the repeated elastic low-energy scattering between the quasifree Rydberg electron at position $\bm{r}$ and the neutral but polarizable ground state atom at position $\bm{R}$ relative to the ionic core.
Fermi's pseudopotential~\cite{Fermi34}, extended to include $p$-wave scattering~\cite{Omont77}, adequately describes this low-energy scattering:
\begin{equation}
\hat{V}_\erAg^{T}(\bm{r}-\bm{R})=2\pi a_0^{T}\delta^3(\bm{r}-\bm{R}) + 6\pi a_1^{T}\delta^3(\bm{r}-\bm{R})\vectLeft{\nabla}\cdot\vectRight{\nabla},
\label{eq:erAgPot}
\end{equation}
where $a_{l}^{T}(k)=-\tan{\left[\delta_{l}^{T}(k)\right]}/k^{2l+1}$ is the energy-de\-pen\-dent triplet ($T$) $s$-wave ($l=0$) and $p$-wave ($l=1$) scattering length, respectively, as a function of the respective phase shifts $\delta_{l}^{T}(k)$~\cite{Supplement}\nocite{Goy86a,Lorenzen83a,Bushaw07a,Marinescu94a,Gallagher94a,Biedenharn81a,Bhatti81a,Sinfailam73a,Norcross71a,Sadeghpour00a,Beckmann74a,Fey15a,Eiles17a,Radziemski95a,Duarte11a,Aymar76a,KnowlesJCP93a,DunningJCP89a,BoysMP70a,ReiherTCA06a,molproa,KokooulineJCP99a,TomzaMP13a,KosloffARPC94a,TomzaPRA12a,qdyna,Osterwalder99a}.
The wave number $k$ of the Rydberg electron (the wave number of the ultracold ground state atom is negligible) at position $\bm{R}$ is given by the semiclassical approximation $k(R)^2/2=1/R-1/[2(\tilde{n}^\effS)^2]$, where $\tilde{n}^\effS$ is the effective principal quantum number of the Rydberg level of interest~\cite{GreenSadeghpour00}.
The Rydberg molecule Hamiltonian, combining all three binary interactions between the Rydberg electron \er, the Rydberg ionic core \Ir, and the ground state atom \Ag, then reads
\begin{equation}
\hat{H}_\ArAg(\bm{r},\bm{R})=\hat{H}_\eIr(\bm{r})+\hat{H}_\IrAg^{\gg}(\bm{R})+\hat{V}_\erAg^{T}(\bm{r}-\bm{R})\hat{P}^{T},
\label{eq:HArAg}
\end{equation}
where $\hat{H}_\eIr$ includes the spin-orbit coupling~\cite{Supplement} and $\hat{H}_\IrAg^{\gg}(\bm{R})=\IrAgPo^2/(2\mu_\IrAg)-C_4/(2R^4)$ describes the \IrAg~interaction at large internuclear distances,
 with $\IrAgPo$ and $\mu_\IrAg$ being the momentum and the reduced mass, respectively.
$\hat{P}^{T}=\erso\cdot\egso+3/4$ is the triplet projection operator, with $\erso$ and $\egso$ the electronic spin of the Rydberg and the ground state electron, respectively.
Applying the Born-Oppenheimer approximation then yields the nuclear Schr\"{o}dinger equation $[\IrAgPo^2/(2\mu_\IrAg)+V_\ArAg(R)]\Psi_\ArAg(\bm{R})=E_\ArAg\Psi_\ArAg(\bm{R})$ with the spherically symmetric Rydberg molecule potentials $V_\ArAg$, the Rydberg molecule wave functions $\Psi_\ArAg$, and the corresponding binding energies $E_\ArAg$ calculated numerically~\cite{Supplement}.
The Rydberg molecule wave function of interest, $\tilde{\Psi}_\ArAg$, is displayed in Fig.~\ref{fig:InitialState}\fsl{a} for both lithium isotopes.
It is bound in the \rt{Li}($30s_{1/2}$)\,-\,\gt{Li}($2s_{1/2}$)~$^3\Sigma$ Rydberg molecule potential $\tilde{V}_\ArAg$, and it is in its rovibrational ground state ($\tilde{v}=\tilde{J}=0$)~\cite{RoVibGround}, thus being spherically symmetric.

For the chosen Rydberg electron principal quantum number $\tilde{n}=30$, the first excited vibrational state ($v=1$) is approximately \SI{5}{MHz} above the ground state, and the first excited rotational state ($J=1$) is approximately $2\tilde{B}\approx\SI{300}{kHz}$ above the ground state [see Fig.~\ref{fig:InitialState}\fsl{b}], where $B=(2\mu_\IrAg r_\textrm{cl}^2)^{-1}$ is the rotational constant of the Rydberg molecule with the internuclear distance being approximated by the classical turning point $r_\textrm{cl}=2(n^\effS)^2$ of the Rydberg electron.
Hence, $2\tilde{B}$ is approximately \num{60} times larger than the typical Rydberg molecule decay linewidth $\gamma_\ArAg$ in a dilute atomic cloud~\cite{Butscher11,Camargo16}.
Besides, broadening mechanisms associated with the photoassociation of the Rydberg molecule (Doppler broadening, power broadening, laser linewidth, and finite laser pulse duration) can be made much smaller than $2\tilde{B}$ in an ultracold atomic cloud and with suitable laser parameters.
Thus, the rovibrational ground state of the Rydberg molecule of interest can be experimentally addressed.
Working with a spin-polarized atomic cloud and appropriate photoassociation laser polarization(s) also justifies neglecting singlet scattering between the Rydberg electron and the ground state atom~\cite{Supplement}.
Furthermore, focusing down the photoassociation laser(s) allows for the formation of a single Rydberg molecule per atomic cloud by means of the Rydberg blockade~\cite{Saffman10}.
In combination with a dilute atomic cloud, this ensures that after photoionization there is only a single ion-atom scattering event happening at a time.

Photoionization frees the initial Rydberg molecule wave function by removing the Rydberg electron.
Choosing a suitable photoionization scheme ensures, first, that the energy imparted onto the ion-atom system in the photoionization process is negligible in comparison with the $S$-wave scattering limit and, second, that the ionization process is fast compared to the effective trapping frequency of the Rydberg molecule potential~\cite{Supplement}.
This makes the photoionization diabatic; i.e., the shape of the initial Rydberg molecule wave function is preserved during ionization.
The initial shape $\Psi_\IrAgS(\bm{R},t=0)$ of the ion-atom wave packet is thus set for the subsequent ultracold scattering.

The scattering of the initial ion-atom wave packet is then described by the time-dependent Schr\"{o}dinger equation
\begin{equation}
i\frac{\partial}{\partial t}\Psi_\IrAgS(\bm{R},t)=\hat{H}_\IrAgM(\bm{R})\Psi_\IrAgS(\bm{R},t),
\label{eq:SEIrAgS}
\end{equation}
where $\hat{H}_\IrAgM$ is the nuclear Hamiltonian of the molecular ion~\cite{Supplement}, containing the spherically symmetric ground state molecular ion potential $\tilde{V}_\IrAgM(R)$.
We carried out \emph{ab initio} calculations to determine $\tilde{V}_\IrAgM$~\cite{Supplement} which yielded a potential depth $D_e=\SI{10468}{cm^{-1}}$ with a conservatively estimated error of \SI{\pm 10}{cm^{-1}}, thus in excellent agreement with the measured value of \SI{10464\pm 6}{cm^{-1}}~\cite{Bernheim84}.
For our molecular ion potential with $D_e=\SI{10468}{cm^{-1}}$, we calculate a $\rtAI{Li}{6}$\,-\,$\gtA{Li}{6}$ ion-atom doublet $S$-wave scattering length of $\mathcal{A}_6=\SI{-1014}{\au}$, with bounds ($\mathcal{A}_6^-;\mathcal{A}_6^+)=(\num{-778};\num{-1294})\,\si{\au}$ corresponding to potentials scaled by $(\num{0.999};\num{1.001})$, respectively, reflecting our accuracy of \SI{0.1}{\percent} in $D_e$.
For $\rtAI{Li}{7}$\,-\,$\gtA{Li}{7}$ we determine the scattering length $\mathcal{A}_7$ to be \SI{7162}{\au}, with bounds of $(\num{107825};\num{3664})\,\si{\au}$
Hence, the scattering lengths differ considerably for the two lithium isotopes, with the magnitude of $\mathcal{A}_6$ being comparable to the characteristic radius of ion-atom interaction $R_6^\effS=\sqrt{\mu_\IrAg C_4}=\SI{949}{\au}$~\cite{Zipkes10a}, whereas $\mathcal{A}_7$ is approximately \num{7} times $R_7^\effS=\SI{1025}{\au}$
This reflects also in the markedly dissimilar bounds we predict for the two scattering lengths, with tight bounds on $\mathcal{A}_6$, while $\mathcal{A}_7$ is extremely sensitive to changes of the molecular ion potential.
Correspondingly, the last bound molecular ion wave function $^7\tilde{\Psi}_\IrAgM$ has a large extension and accordingly a small binding energy, whereas $^6\tilde{\Psi}_\IrAgM$ is deeply bound at comparatively small internuclear distances, as can be seen in Fig.~\ref{fig:InitialState}\fsl{a} (note the logarithmic scale for the internuclear distance).
The vibrational molecular ion wave functions $\Psi_\IrAgM^v(\bm{R})$ and corresponding eigenenergies $E_\IrAgM^v$, with $\hat{H}_\IrAgM(\bm{R})\Psi_\IrAgM^v(\bm{R})=E_\IrAgM^v\Psi_\IrAgM^v(\bm{R})$, are then used to express the scattered ion-atom wave packet in the form
\begin{equation}
\Psi_\IrAgS(\bm{R},t)=\sum_{v}{\pi_v\,e^{-i E_\IrAgM^v t}\,\Psi_\IrAgM^v(\bm{R})},
\label{eq:PsiIrAgS}
\end{equation}
where the sum runs over bound ($E_\IrAgM^v<0$) and scattering states ($E_\IrAgM^v>0$) and $\pi_v=\Braket{\Psi_\IrAgS(\bm{R},t=0)|\Psi_\IrAgM^v(\bm{R})}$ is the projection of the initial ion-atom wave packet on the vibrational state $\Psi_\IrAgM^v$.
With the ion-atom interaction potential $\tilde{V}_\IrAgM(R)$ being spherically symmetric, and without an electric field $\mathcal{E}$ present, the initial orbital angular momentum $J$ of the ion-atom wave packet, given by the Rydberg molecule wave function, is conserved during scattering; i.e., scattering channels for different partial waves do not couple.
Consequently, the molecular ion wave functions used to express the scattered wave packet in Eq.~\ref{eq:PsiIrAgS} have this very orbital angular momentum $J$.
For the case studied in this Letter, where the initial Rydberg molecule wave function is in its spherically symmetric rotational ground state ($\tilde{J}=0$), the orbital angular momentum conservation implies that only $S$-wave ion-atom scattering can occur despite the fact that $\tilde{\Psi}_\ArAg$ has an energy spectrum with components above the $S$-wave scattering limit, as is illustrated in Fig.~\ref{fig:InitialState}\fsl{c}.
The effect of electric stray fields $\mathcal{E}_\textrm{stray}$ is discussed in Supplemental Material~\cite{Supplement}.

\begin{figure}[htbp]
\centering
\includegraphics[width=1.01\columnwidth]{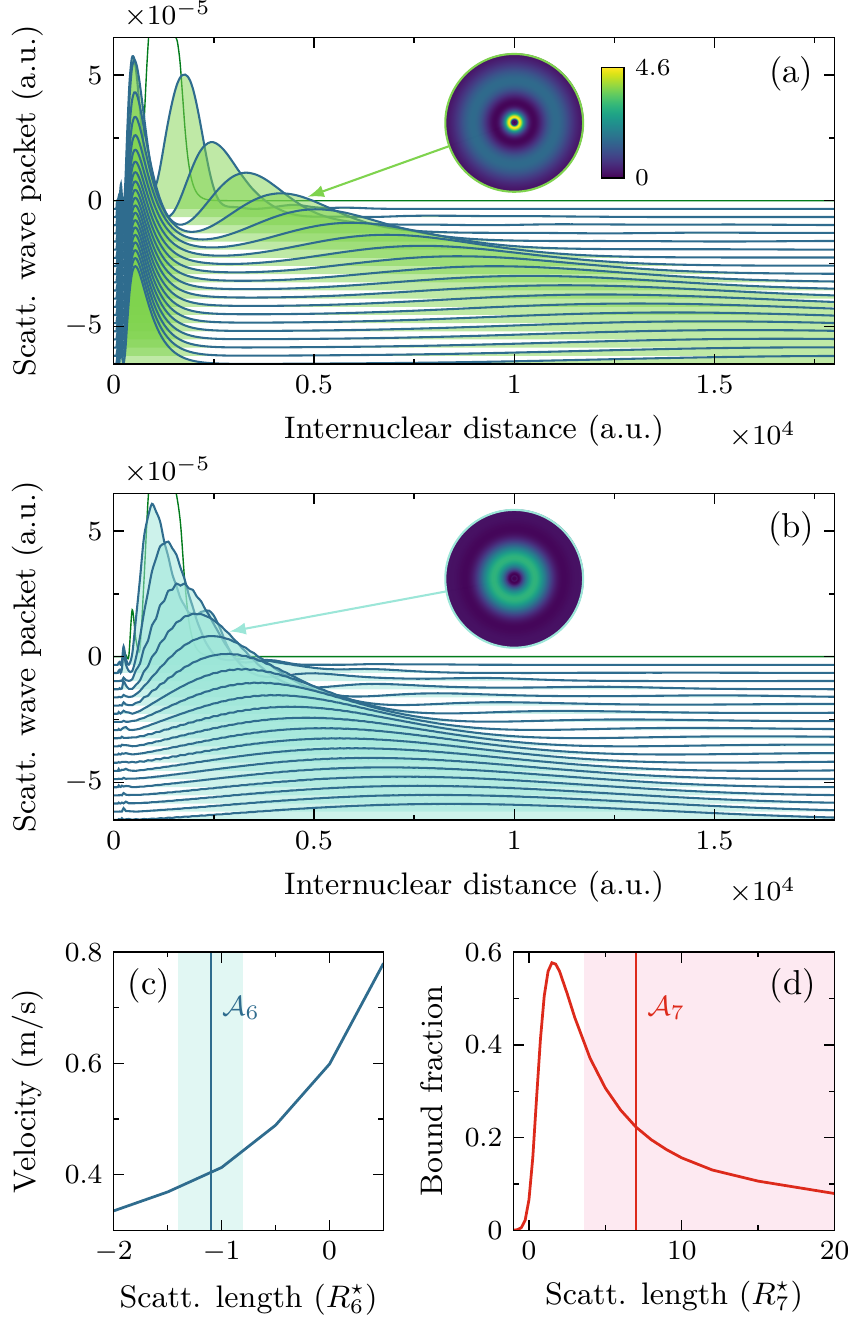}
\vspace{-0.45cm}
\caption{
Ultracold lithium ion-atom scattering processes, initialized with Rydberg molecules, for different scattering lengths $\mathcal{A}$.
For $\mathcal{A}>0$ [demonstrated in \fsl{a} for $\rtAI{Li}{6}$\,-\,$\gtA{Li}{6}$ scattering with $\mathcal{A}=+R_6^\effS$], the scattered wave packet splits into a free, expanding shell and a bound shell, indicating molecular ion formation.
For $\mathcal{A}<0$ [exemplified in \fsl{b} for $\mathcal{A}=-R_6^\effS$], the entire scattered wave packet is free.
In \fsl{a} and \fsl{b}, the time evolution of the scattered wave packet $R^2\left|\Psi_\IrAgS(R,t)\right|^2$ is shown from $t=0$ [$\Psi_\IrAgS(t=0)=\tilde{\Psi}_\ArAg$, unfilled curve, vertical axis applies] to \SI{1}{\mu s} in steps of \SI{50}{ns} (top to bottom, curves shifted by \SI{-3.25e-6}{\au} each).
The insets demonstrate the $S$-wave character of the scattered wave packet [$R$ extension of \SI{7e3}{\au}, same color scale in \fsl{a} and \fsl{b}].
For $\mathcal{A}<0$ ($\mathcal{A}>0$), the shell expansion velocity $\zeta$ (the bound fraction $b$) is a sensitive quantity to determine the scattering length; see \fsl{c} and \fsl{d}, where also the scattering lengths $\mathcal{A}_{6,7}$ from our \emph{ab initio} calculations are indicated (vertical lines with the shaded areas marking the bounds).
}
\label{fig:ScatteredWavepacket}%
\end{figure}

We conducted ion-atom scattering calculations for both lithium isotopes, with $\Psi_\IrAgS(\bm{R},t=0)$ given by the respective Rydberg molecule wave function $\tilde{\Psi}_\ArAg$ displayed in Fig.~\ref{fig:InitialState}\fsl{a}, and for the respective ion-atom scattering length tuned over a wide range~\cite{Supplement}.
These calculations revealed two different scattering regimes depending on the scattering length $\mathcal{A}$.
For positive scattering lengths, exemplarily illustrated in Fig.~\ref{fig:ScatteredWavepacket}\fsl{a} for $\rtAI{Li}{6}$\,-\,$\gtA{Li}{6}$ scattering with a scattering length of $+R_6^\effS$, the scattered wave packet splits into a free, dispersively expanding shell and a bound shell, the position and shape of which reveal that the last bound molecular ion state has been formed.
Both shells are spherically symmetric and thus demonstrate that only $S$-wave scattering occurs.
For negative scattering lengths, the entire scattered wave packet is free as is exemplarily shown in Fig.~\ref{fig:ScatteredWavepacket}\fsl{b} for $\rtAI{Li}{6}$\,-\,$\gtA{Li}{6}$ scattering with a scattering length of $-R_6^\effS$.
This is due to the negligible overlap between the initial Rydberg molecule wave function and the last bound molecular ion wave function, as can be seen in Fig.~\ref{fig:InitialState}\fsl{a} for the example of $^\textrm{6}\tilde{\Psi}_\ArAg$ and $^6\tilde{\Psi}_\IrAgM$ with $\mathcal{A}_6=\num{-1.1}R_6^\effS$.
In the regime of negative scattering lengths, the velocity $\zeta$ with which the maximum of the expanding shell moves is a sensitive quantity to determine the scattering length, as can be seen from Fig.~\ref{fig:ScatteredWavepacket}\fsl{c}.
With our predicted bounds on $\mathcal{A}_6$, the $\rtAI{Li}{6}$\,-\,$\gtA{Li}{6}$ scattering falls into this regime.
For positive scattering lengths, the bound fraction $b$ of the total scattered ion-atom wave packet can be used to precisely determine the scattering length, as is demonstrated in Fig.~\ref{fig:ScatteredWavepacket}\fsl{d} for $\rtAI{Li}{7}$\,-\,$\gtA{Li}{7}$ scattering, falling in this regime according to our scattering length calculations.

To experimentally measure the expansion velocity $\zeta$, the freely moving scattered ion of a single ultracold scattering event is imaged onto a time- and position-sensitive single-ion detector.
After many repetitions of this single scattering event, the scattered ion-atom wave packet can eventually be reconstructed, either in momentum space when, e.g., using a \textsc{motrims}~\cite{Goetz12} or \textsc{vmi}~\cite{Whitaker03} apparatus or in real space when employing an ion microscope for imaging.
Given the submicron resolution of these ion microscopes~\cite{Schwarzkopf13,Stecker17}, an evolution time in the microsecond range is sufficient to resolve the shape of the scattered wave packet.
The bound fraction $b$ can be measured in time of flight where the molecular ions $\rtMI{Li}$ separate from the lighter free $\rtI{Li}$ ions.
For guiding the ions onto the detector, a suitably small extraction electric field has to be used not to dissociate the weakly bound molecular ions.

The presented experimental method to use Rydberg molecules for the investigation of ultracold ion-atom scattering, particularized for the lithium ion-atom system in this Letter, is readily applicable to other ion-atom systems, e.g., to homo- or heteronuclear alkali or alkaline earth ion-atom systems for which Rydberg molecules can be formed~\cite{Sinfailam73a,Bahrim01,Bartschat03}.
In such a manner, the \emph{ab initio} calculations of ion-atom scattering lengths could be benchmarked for increasingly complicated systems.
Furthermore, having the ion-atom scattering lengths precisely determined with the proposed two-body scattering experiment would immediately allow for a more faithful and accurate description of the many-body, polaronic properties of an ion impurity immersed in an atomic quantum gas~\cite{Cote02,Massignan05,Casteels11,Schurer14,Schurer15}.
Finally, keeping the Rydberg electron as a spectator and Faraday cage for the ion-atom collision~\cite{Pratt94,Strazisar01,Wrede05,Dai05,Allmendinger16} might be possible also in the ultracold regime using circular Rydberg states~\cite{Matsuzawa10}.

\begin{acknowledgments}
We thank K. Jachymski and F. Merkt for fruitful discussions.
We thank S. Hofferberth, K. Kleinbach, and F. Meinert for numerical code development to calculate Rydberg molecule potentials and wave functions.
We acknowledge support from Deutsche Forschungsgemeinschaft [Projects No. PF 381/13-1 and No. PF 381/17-1, the latter being part of the SPP 1929 (GiRyd)], from the Czech Science Foundation (Project No. P208/14-15989P), and from the National Science Centre Poland (Project No. 2016/23/B/ST4/03231) as well as from the PL-Grid Infrastructure.
C.\,V. acknowledges support from the Carl-Zeiss-Stiftung.
\end{acknowledgments}
\parbox[c][0.5cm][c]{\columnwidth}{}

\makeatletter
\interlinepenalty=10000
\putbib[ias-ars]
\makeatother
\end{bibunit}

\clearpage
\renewcommand{\thefigure}{\SupplementPrefix\arabic{figure}}
\setcounter{figure}{0}
\renewcommand{\thetable}{\SupplementPrefix\Roman{table}}
\setcounter{table}{0}
\renewcommand{\theequation}{\SupplementPrefix\arabic{equation}}
\setcounter{equation}{0}
\renewcommand{\citenumfont}[1]{S#1}
\renewcommand{\bibnumfmt}[1]{[S#1]}

\section{Supplemental Material \lowercase{for}: ``R\lowercase{ydberg} M\lowercase{olecules for} I\lowercase{on-}A\lowercase{tom} S\lowercase{cattering in the} U\lowercase{ltracold} R\lowercase{egime}''}
\label{sec:Supplemental}
\setcounter{page}{1}
\begin{bibunit}

\subsection{Rydberg molecule calculations}
\label{subsec:SupplementRydbergMoleculeCalc}
\begin{figure}[htbp]
\centering
\includegraphics[width=0.41\columnwidth]{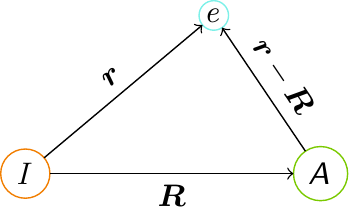}
\caption{
Schematic illustration of the Rydberg molecule~\ArAg, constituting a three-body system where the Rydberg electron~\er\ binds the neutral, polarizable ground state atom \Ag\ to the Rydberg ionic core \Ir.
}
\label{fig:3bodySystems}
\end{figure}

In the following, the calculation of Rydberg molecule wave functions and binding energies is thoroughly discussed.
First, two of the three binary interactions of the three-body Rydberg molecule Hamiltonian (see Eq.~\ref{eq:HArAg} and Fig.~\ref{fig:3bodySystems}) are detailed, namely the Rydberg-electron -- Rydberg-ionic-core interaction and the Rydberg-electron -- ground-state-atom interaction.
Afterwards, the solution of the complete Rydberg molecule Hamiltonian by applying the Born-Oppenheimer approximation, i.e., by treating the electronic and nuclear degrees of freedom sequentially, is particularized.

\subsubsection{Rydberg-electron -- Rydberg-ionic-core interaction}
\label{subsubsec:Supplement-erIr}
The Schr{\"o}dinger equation for the Rydberg-electron -- Rydberg-ionic-core interaction reads
\begin{equation}
\left[\frac{\hat{\bm{p}}^2}{2\mu_{\eIr}} + \hat{V}_{\eIr}(r)\right]\psi_{nljm_j}(\bm{r})=E_{nlj}\psi_{nljm_j}(\bm{r}),
\label{eq:SEpsinljmj}
\end{equation}
where $\hat{\bm{p}}$ is the Rydberg electron momentum, $\mu_{\eIr}$ is the reduced electron -- ionic-core mass, which is approximated with the bare electron mass $m_e$ in the following, $\hat{V}_{\eIr}(r)$ is the spherically symmetric interaction potential, $\psi_{nljm_j}$ are the Rydberg electron wave functions, and $E_{nlj}$ are the corresponding eigenenergies.

For $l\leq 2$, the eigenenergies $E_{nlj}$ are calculated via
\begin{equation}
E_{nlj}=-\frac{\mathcal{R}^\effS}{\left(n^\effS\right)^2},
\label{eq:Enljlowl}
\end{equation}
where $n^\effS$ is the effective principal quantum number given by
\begin{equation}
n^\effS=n-\delta_{nlj},
\label{eq:nStar}
\end{equation}
with the quantum defect
\begin{equation}
\delta_{nlj}=\delta_0+\frac{\delta_2}{(n-\delta_0)^2}+\frac{\delta_4}{(n-\delta_0)^4}+\frac{\delta_6}{(n-\delta_0)^6}+\dots
\label{eq:QuantumDefect}
\end{equation}
(see Table~\ref{tab:LiQuantumDefects} for the used values for lithium), and where $\mathcal{R}^\effS$ is the reduced Rydberg constant (given in Table~\ref{tab:LiVariousConstants} for lithium).
\begin{table*}[htbp]%
\caption{Lithium quantum defects used in this Letter.}
\label{tab:LiQuantumDefects}
\begin{tabular}{
l
l
S[table-format=1.7]
S[table-format=-1.5]
S[table-format=1.4]
S[table-format=-1.4]
r}
\toprule
Isotope & State & {$\delta_0$} & {$\delta_2$} & {$\delta_4$} & {$\delta_6$} & Ref. \\
\midrule
6,\,7 & $s_{1/2}$ & 0.3995101 & 0.0290 & & & \cite{Goy86b} \\
6 & $p_{1/2}$ & 0.0471835 & -0.024 & & & \cite{Goy86b} \\
6 & $p_{3/2}$ & 0.0471720 & -0.024 & & & \cite{Goy86b} \\
7 & $p_{1/2}$ & 0.0471780 & -0.024 & & & \cite{Goy86b} \\
7 & $p_{3/2}$ & 0.0471665 & -0.024 & & & \cite{Goy86b} \\
6,\,7 & $d_{3/2}$,\,$d_{5/2}$ & 0.002129 & -0.01491 & 0.1759 & -0.8507 & \cite{Lorenzen83b} \\
\bottomrule
\end{tabular}
\end{table*}
\begin{table}[htbp]%
\caption{Lithium reduced Rydberg constants $\mathcal{R}^\effS$ and lithium ionic core polarizability $\alpha_c$ used in this Letter.}
\label{tab:LiVariousConstants}
\begin{tabular}{
c
l
S[table-format=4.8]
r}
\toprule
Quantity & Isotope & {Value} & Ref. \\
\midrule
$\mathcal{R}^\effS$ (THz) & 6 & 3289.541926 & \cite{Goy86b} \\
$\mathcal{R}^\effS$ (THz) & 7 & 3289.58472832 & \cite{Bushaw07b} \\
$\alpha_c$ (\si{\au}) & 6,\,7 & 0.1923 & \cite{Marinescu94b} \\
\bottomrule
\end{tabular}
\end{table}
For $l\geq 3$, the eigenenergies are calculated with~\cite{Gallagher94b}
\begin{equation}
E_{nlj}=-\mathcal{R}^\effS\left[\frac{1}{n^2}+\frac{\alpha_{\text{fs}}^2}{n^3}\left(\frac{1}{j+1/2}-\frac{3}{4n}\right)\right]-\frac{3\alpha_c}{4n^3l^5},
\label{eq:Enljhighl}
\end{equation}
where $\alpha_{\text{fs}}$ is the fine structure constant, and $\alpha_c$ is the ionic core polarizability (see Table~\ref{tab:LiVariousConstants} for the lithium value).

For $\hat{V}_{\eIr}(r)$, the following model potential is used~\cite{Marinescu94b}:
\begin{equation}
\hat{V}_{\text{\eIr}}(r)=V_C(r) + V_p(r) + \hat{V}_{\text{so}}(r),
\label{eq:VeIr}
\end{equation}
with the Coulomb term
\begin{equation}
V_C(r)=-\frac{Z_l(r)}{r},
\label{eq:VC}
\end{equation}
$Z_l$ being the radial charge distribution
\begin{equation}
Z_l(r)=1+(Z-1)\text{e}^{-\alpha_1 r} - r(\alpha_3 + \alpha_4 r)\text{e}^{-\alpha_2 r},
\label{eq:Zl}
\end{equation}
with the nuclear charge $Z$ and $\alpha_{1,2,3,4}$ model parameters (see Ref.~\cite{Marinescu94b} for lithium values);
$V_p$ is the core polarization term
\begin{equation}
V_p(r)=-\frac{\alpha_c}{2r^4}\left[1-e^{-\left(r/r_c\right)^6}\right],
\label{eq:Vp}
\end{equation}
with $r_c$ the effective core size (see Ref.~\cite{Marinescu94b});
$\hat{V}_{\text{so}}$ is the spin-orbit coupling term for which the approximation
\begin{equation}
\hat{V}_{\text{so}}(r>r_c)=\frac{\alpha_{\text{fs}}^2}{2r^3}\hat{\bm{l}}\cdot\erso
\label{eq:VsoOp}
\end{equation}
is used, valid in the region $r>r_c$.
For calculating Rydberg wave functions, this approximation is justified.

To solve the Schr\"{o}dinger equation~\ref{eq:SEpsinljmj} for the Rydberg electron wave functions, the separation ansatz
\begin{equation}
\psi_{nljm_j}(\bm{r})=R_{nlj}(r)\mathcal{Y}_{(l)jm_j}(\vartheta,\varphi)
\label{eq:psinljmjsep}
\end{equation}
is made, where $\mathcal{Y}_{(l)jm_j}$ are the spin spherical harmonics for $s=1/2$, thus $l=j\pm1/2$, given in matrix form by~\cite{Biedenharn81b}
\begin{align}
\mathcal{Y}_{(j\pm\frac{1}{2})jm_j}=&\frac{1}{\sqrt{2(j\pm\frac{1}{2})+1}} \nonumber \\
&\times\begin{pmatrix}\mp\sqrt{j\pm\frac{1}{2}\mp m_j +\frac{1}{2}}Y_{j\pm\frac{1}{2}}^{m_j-\frac{1}{2}}\\ \sqrt{j\pm\frac{1}{2}\pm m_j +\frac{1}{2}}Y_{j\pm\frac{1}{2}}^{m_j+\frac{1}{2}}\end{pmatrix},
\label{eq:spinY}
\end{align}
with $Y_{l}^{m_l}$ being the spherical harmonics.
The spin-orbit term acting on this ansatz yields
\begin{align}
\hat{V}_{\text{so}}\psi_{nljm_j}&=\frac{\alpha_{\text{fs}}^2}{4r^3}\left[j(j+1)-l(l+1)-\frac{3}{4}\right]\psi_{nljm_j} \nonumber \\
&=\mathcal{Y}_{(l)jm_j}V_{\text{so}}R_{nlj},
\label{eq:VsoSpinY}
\end{align}
where
\begin{equation}
V_{\text{so}}(r)=\frac{\alpha_{\text{fs}}^2}{4r^3}\left[j(j+1)-l(l+1)-s(s+1)\right]
\label{eq:Vso}
\end{equation}
was introduced with $s=1/2$.
The radial part $R_{nlj}$ of the electron wave functions is then found solving
\begin{equation}
\left[-\frac{1}{2r}\frac{\text{d}^2}{\text{d}r^2}r + \frac{l(l+1)}{2r^2} + V_{\eIr}(r)\right]R_{nlj}(r)=E_{nlj}R_{nlj}(r)
\label{eq:SE-Rnlj}
\end{equation}
with a square root rescaling of $r$~\cite{Bhatti81b} and by using Numerov's method for numerical integration.

\subsubsection{Rydberg-electron -- ground-state-atom interaction}
\label{subsubsec:Supplement-erAg}
\begin{figure}[htbp]
\centering
\includegraphics[width=0.988\columnwidth]{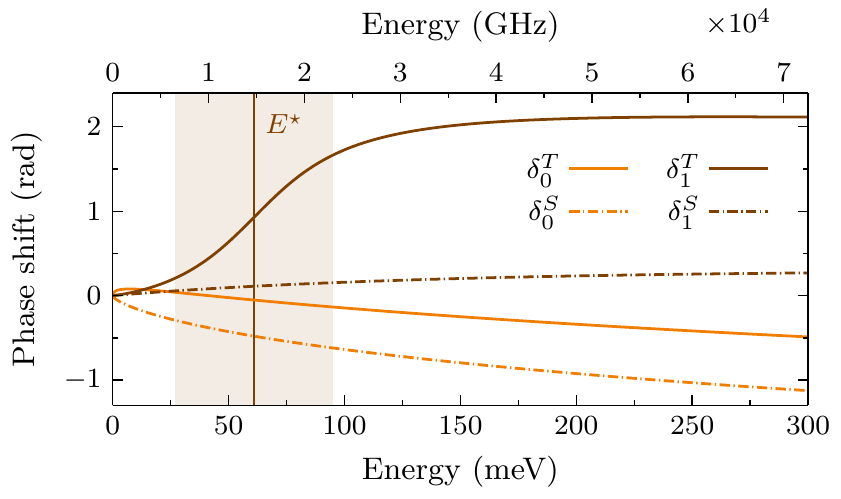}
\caption{
Non-isotope-specific electron-lithium scattering phase shifts $\delta^{S,T}_l$ as a function of energy $E$ for singlet ($S$) and triplet ($T$), $s$-wave ($l=0$) and $p$-wave ($l=1$) scattering.
The triplet $p$-wave shape resonance position $E^\effS$ and width $\Gamma^\effS$ (shaded area) are also indicated (the values are given in Table~\ref{tab:eLiScattPar}).
The presented phase shifts feature a high accuracy at low scattering energies.
}
\label{fig:ElectronLiPhaseshifts}%
\end{figure}

As outlined in the main article, the Rydberg-electron -- ground-state-atom interaction potential is given by Eq.~\ref{eq:erAgPot}, where the energy-dependent triplet $s$- and $p$-wave phase shifts $\delta_{l}^{T}(k)$ are used to calculate the respective scattering lengths $a_{l}^{T}(k)$.
For the lithium system used in this Letter to exemplify the proposed method for ultracold ion-atom scattering, we conducted $R$-matrix calculations to determine the electron-lithium scattering phase shifts.
These phase shifts are displayed in Fig.~\ref{fig:ElectronLiPhaseshifts}, and characteristic quantities extracted from them are given in Table~\ref{tab:eLiScattPar}.\nocite{Sinfailam73b,Norcross71b}
Our phase shifts feature a better accuracy than previously published values (see Table~\ref{tab:eLiScattPar}), especially at very low scattering energies, which is necessary to faithfully represent the low-energy scattering between the Rydberg electron and the ground state atom.

For internuclear distances larger than the classical turning point of the Rydberg electron, the semiclassical approximation
\begin{equation}
\frac{k^{2}}{2}=\frac{1}{R}-\frac{1}{2\left(n^\effS\right)^2}
\label{eq:kSemiClassical}
\end{equation}
for the scattering energy $E=k^{2}/2$ ceases to be applicable.
Additionally, for the polarization potential, the $p$-wave scattering lengths diverge in the zero-energy limit~\cite{Sadeghpour00b}.
Therefore, a cutoff energy is defined below which, in Eq.~\ref{eq:erAgPot}, the zero-energy $s$-wave scattering lengths are used and the $p$-wave scattering lengths are set to zero.
This cutoff energy is defined such that the Rydberg molecule potential of interest $\tilde{V}_\ArAg$ is smooth at the classical turning point.
It is approximately \SI{0.276}{meV} for the $\tilde{V}_\ArAg$ shown in Fig.~\ref{fig:InitialState}\fsl{a}.

In this Letter, singlet scattering between the Rydberg electron and the ground state atom is neglected as well as the hyperfine interaction in the ground state atom.
This yields a prototypical Rydberg molecule as for example in Ref.~\cite{BendkowskyPfau10b} which is, for its simplicity, well suited to demonstrate the proposed method for ultracold ion-atom scattering.
Conducting the experiment with stretched states of the Rydberg and the ground state electron (e.g., $m_s=\eIgVi{m}{s}=+1/2$), and that the depth of the Rydberg molecule potential of interest in the range given by the extension of $\tilde{\Psi}_\ArAg$ is much smaller than the hyperfine splitting in the lithium ground state atom \{approximately \SI{228}{MHz} for $\gtA{Li}{6}$ and \SI{804}{MHz} for $\gtA{Li}{7}$~\cite{Beckmann74b} in comparison with about \SI{30}{MHz}; see Fig.~\ref{fig:InitialState}\fsl{a}\} guarantees that these two omissions are justified.

\begin{table}[htbp]%
\caption{
Electron-lithium scattering: characteristic quantities and accuracy considerations.
Given are the zero-energy singlet ($S$) and triplet ($T$) $s$-wave ($l=0$) scattering lengths $\bar{a}_0^{S,T}$ and the triplet $p$-wave shape resonance position $E^\effS$ and width $\Gamma^\effS$.
Our values (PW: present work) are in good agreement with previously published ones, except for $\Gamma^\effS$ where the deviation might be attributed to convergence issues in the calculations of Ref.~\cite{Sinfailam73b}.
Comparing the minimum scattering energies $E_\text{min}$ down to which the respective phase shift calculations were conducted and the results tabulated (with the energy resolution being on the same order as $E_\text{min}$) indicates the improved accuracy of our calculations, especially at low scattering energies.
This, in turn, renders our zero-energy scattering lengths particularly sound.}
\label{tab:eLiScattPar}
\begin{tabular}{
S[table-format=-1.2]
S[table-format=1.2]
S[table-format=2.1]
S[table-format=2.1]
S[table-format=2.3]
c}
\toprule
{$\bar{a}_0^{T}$ (\si{\au})} & {$\bar{a}_0^{S}$ (\si{\au})} & {$E^\effS$ (meV)} & {$\Gamma^\effS$ (meV)} & {$E_\text{min}$ (meV)} & Ref. \\
\midrule
-7.43 & 2.99 & 60.9 & 67.9 & 0.023 & PW\\
-7.12 & 3.04 & & & 0.136 & \cite{Norcross71b} \\
 & & 60 & 57 & 34 & \cite{Sinfailam73b} \\
\bottomrule
\end{tabular}
\end{table}

\subsubsection{Solving the electronic Schr\"{o}dinger equation}
\label{subsubsec:Supplement-ArAgElectronic}
Treating the nuclear coordinates as parameters, the electronic Schr\"{o}dinger equation of the Rydberg molecule reads
\begin{equation}
\hat{H}_\ArAg(\bm{r};R)\psi_\ArAg(\bm{r};R)=V_\ArAg(R)\psi_\ArAg(\bm{r};R),
\label{eq:eSEArAg}
\end{equation}
where
\begin{equation}
\hat{H}_\ArAg(\bm{r};R)=\hat{H}_\eIr(\bm{r})+V_\IrAg^{\gg}(R)+\hat{V}_\erAg^{T}(\bm{r}-R\bm{e}_z)\hat{P}^{T}
\label{eq:eHArAg}
\end{equation}
is the electronic Hamiltonian with $\hat{H}_\eIr(\bm{r})$ given in Eq.~\ref{eq:SEpsinljmj}, with $V_\IrAg^{\gg}(R)=-C_4/(2R^4)$ as the ion-atom interaction potential valid at large internuclear distances, and with $\hat{V}_\erAg^{T}(\bm{r}-R\bm{e}_z)$ given by Eq.~\ref{eq:erAgPot}, where $\bm{R}$ was chosen to lie along the $z$ axis, $\bm{R}=R\bm{e}_z$, without loss of generality.
To solve for the electronic Rydberg molecule orbitals (MO) $\psi_\ArAg(\bm{r};R)$ and the corresponding spherically symmetric Rydberg molecule potentials $V_\ArAg(R)$, the Schr\"{o}dinger equation~\ref{eq:eSEArAg} is numerically diagonalized for each $R$ in the basis $\left\{\Ket{n,l,j,m_j;\eIgVi{m}{s}}\right\}$ of Rydberg electron wave functions $\psi_{nljm_j}(\bm{r})$ (see Eq.~\ref{eq:psinljmjsep}) multiplied with the spin state $\Ket{\eIgVi{m}{s}}$ of the ground state atom's valence electron ($\eIgVi{m}{s}=\pm 1/2$ for lithium).
The infinite number of basis states of different $n$ is reduced to comprise four total manifolds, two below and one above the manifold containing the Rydberg level of interest; i.e., for lithium with zero integer parts of the quantum defects (see Table~\ref{tab:LiQuantumDefects}) $n=\tilde{n}-2,\tilde{n}-1,\tilde{n},\tilde{n}+1$, with $\tilde{n}$ indicating the principal quantum number of the Rydberg level of interest.
This was shown to yield accurate Rydberg molecule potentials~\cite{Fey15b,Eiles17b}.
The orbital angular quantum number $l$ is not truncated, thus $l=0,\ldots,n-1$.
With merely $s$- and $p$-wave scattering included (see Eq.~\ref{eq:erAgPot}), only basis states with $\left|m_j\right|\leq 3/2$ contribute since the spherical harmonics vanish on the $z$ axis except for $m_l=0$.
Obtaining the same number of electronic Rydberg MO's $\psi_\ArAg$ and corresponding potentials $V_\ArAg$ as the number of basis states included in the diagonalization, a counting index $N$ is introduced for unambiguous identification, with $\tilde{N}=0$ for the MO and potential of interest, i.e., $\tilde{\psi}_\ArAg=\psi_0$ and $\tilde{V}_\ArAg=V_0$.
Alternatively, the MO and potential of interest are labeled with the term symbol $30s_{1/2}\sigma\,2s_{1/2}\sigma~^3\Sigma_0$.

\subsubsection{Solving the nuclear Schr\"{o}dinger equation}
\label{subsubsec:Supplement-ArAgNuclear}
The Rydberg molecule potentials obtained by solving the electronic Schr\"{o}\-ding\-er equation~\ref{eq:eSEArAg} enter the nuclear Schr\"{o}dinger equation of the Rydberg molecule in the form
\begin{equation}
\left[\frac{\IrAgPo^2}{2\mu_\IrAg}+V_N(R)\right]\Psi_\ArAgNvJ(\bm{R})=E_\ArAgNvJ\Psi_\ArAgNvJ(\bm{R}),
\label{eq:nSEArAg}
\end{equation}
where $\Psi_\ArAgNvJ(\bm{R})$ is the nuclear Rydberg MO, also denoted Rydberg molecule wave function, bound in the $N$'th Rydberg molecule potential, with $v$ and $J$ indicating the vibrational and rotational state, respectively; $E_\ArAgNvJ$ is the corresponding Rydberg molecule binding energy.
Making the separation ansatz
\begin{equation}
\Psi_{\ArAgNvJ}(\bm{R})=R_{\ArAgNvJ}(R)\mathcal{P}_{J}(\theta)
\label{eq:PsiArAgSep}
\end{equation}
with the normalized Legendre polynomials
\begin{equation}
\mathcal{P}_{J}(\theta)=\sqrt{\frac{2J+1}{4\pi}}P_{J}(\theta)
\label{eq:PJArAgNorm}
\end{equation}
for the angular part then yields the one-dimensional radial Schr\"{o}dinger equation
\begin{align}
&\left[-\frac{1}{2\mu_\IrAg}\frac{1}{R}\frac{\text{d}^2}{\text{d}R^2}R+\frac{J\left(J+1\right)}{2\mu_\IrAg R^2}+V_{N}(R)\right]R_{\ArAgNvJ}(R)\nonumber\\
&=E_{\ArAgNvJ}R_{\ArAgNvJ}(R),
\label{eq:n1dSEArAg}
\end{align}
which is solved for the radial part $R_{\ArAgNvJ}(R)$ and binding energy $E_{\ArAgNvJ}$ by applying a square root rescaling of $R$~\cite{Bhatti81b} and subsequent numerical integration using Numerov's method.

\subsection{Photoionization}
\label{subsec:SupplementPhotoionization}
\begin{figure}[htbp]
\centering
\includegraphics[width=0.785\columnwidth]{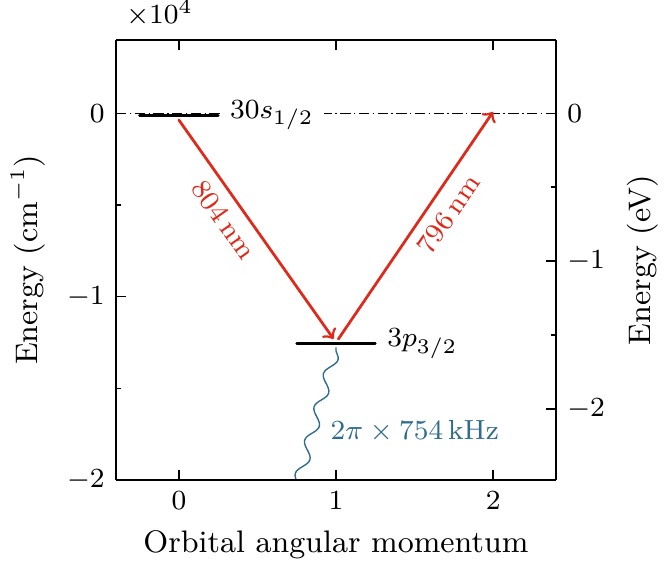}
\caption{
$V$-type photoionization scheme to remove the Rydberg electron of the initial Rydberg molecule in order to start the ultracold ion-atom scattering.
Here, this scheme is illustrated for lithium, with the Rydberg electron initially in the $30s_{1/2}$ state and subsequently brought into the continuum with two lasers via the $3p_{3/2}$ state (the dash-dotted line indicates the ionization threshold).
The energy $E_{30s_{1/2}}$ is calculated with Eq.~\ref{eq:Enljlowl}, $E_{3p_{3/2}}$ is determined from references~\cite{Radziemski95b,Bushaw07b}, and the total decay rate of the $3p_{3/2}$ state (both into the $3s_{1/2}$ and the $2s_{1/2}$ state) is taken from Ref.~\cite{Duarte11b}.
}
\label{fig:Photoionization}%
\end{figure}

We propose a $V$-type photoionization scheme to free the initial Rydberg molecule wave function in order to start the ultracold ion-atom scattering.
Applying this scheme as displayed in Fig.~\ref{fig:Photoionization} to the lithium Rydberg molecule wave function of interest $\tilde{\Psi}_\ArAg$, with the two ionization laser beams copropagating and with the ascending laser tuned maximally \SI{1}{GHz} above the ionization threshold, the energy imparted onto the ion-atom system is more than \num{10} times smaller than the respective $S$-wave scattering limit of the two lithium isotopes.
The photoionization process is diabatic as long as its timescale is considerably faster than the trapping frequencies of the Rydberg molecule potential in which $\tilde{\Psi}_\ArAg$ is bound.
For the photoionization process depicted in Fig.~\ref{fig:Photoionization}, with an ionization cross section taken from Ref.~\cite{Aymar76b} and for typical ionization laser parameters (i.e., for typical laser powers and beam waists), we calculate a photoionization timescale of a few nanoseconds, corresponding to a frequency of a few tens of megahertz.
It is thus significantly larger than the effective radial trapping frequency one can attribute to the Rydberg molecule potential of interest $\tilde{V}_\ArAg$ by averaging the zero-point energy over the main extension range of $\tilde{\Psi}_\ArAg$ [see Fig.~\ref{fig:InitialState}\fsl{a}].
In contrast to the radial direction, there is no angular confinement present for $\tilde{\Psi}_\ArAg$.
Consequently, the photoionization process of the lithium initial Rydberg molecule wave function is diabatic.

\subsection{Molecular ion calculations}
\label{subsec:SupplementMolecularIonCalc}
\begin{figure}[htbp]
\centering
\includegraphics[width=0.523\columnwidth]{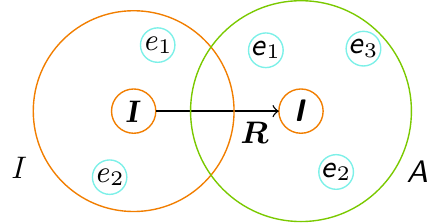}
\caption{
Schematic illustration of the lithium molecular ion~\IrAgM\ studied in this Letter, constituting a seven-body system where in total five electrons (two electrons $\eri{i}$ from the Rydberg ionic core \Ir~and three electrons $\egi{j}$ from the ground state atom \Ag) bind the two triply-charged bare ionic cores \IrB\ and \IgB\ together.
}
\label{fig:IrAgM}
\end{figure}

In the following, our \emph{ab initio} calculations of the lithium ground state molecular ion potential $\tilde{V}_\IrAgM$, also denoted as ion-atom interaction potential, are detailed.
We treat the lithium molecular ion as a seven-body system where the altogether five electrons bind the two triply-charged bare ionic cores together (see Fig.~\ref{fig:IrAgM}).
As opposed to the three-body Rydberg molecule system (see Fig.~\ref{fig:3bodySystems}), where the Rydberg electron is singled out and the remaining electrons enter the Rydberg molecule calculations only implicitly, for our molecular ion calculations, all electrons are treated on equal footing.

We apply the Born-Oppenheimer approximation to our five-electron -- two-nuclei lithium molecular ion system in order to separate the electronic from the nuclear degrees of freedom.
This yields the electronic Schr\"{o}dinger equation, the solution of which in turn yields the spherically symmetric molecular ion potentials $V_\IrAgM(R)$.
For our ion-atom scattering calculations, only the ground state molecular ion potential $\tilde{V}_\IrAgM(R)$ is of interest (see Eq.~\ref{eq:nHIrAgM}).
To solve the electronic Schr\"{o}dinger equation for $\tilde{V}_\IrAgM$, we use coupled cluster methods with Gaussian basis sets.
In the first step, we use the spin-restricted open-shell coupled cluster method restricted to single, double, and non-iterative triple excitations, starting from the restricted open-shell Hartree-Fock orbitals, RCCSD(T)~\cite{KnowlesJCP93b}, and the augmented correlation-consistent polarized core-valence quintuple-\textzeta\ quality basis set (aug-cc-pCV5Z)~\cite{DunningJCP89b}. 
The interaction energy is obtained with the supermolecule method and the basis set superposition error is corrected by using the counterpoise correction~\cite{BoysMP70b}.
Next, the remaining contribution of the full triple excitations in the coupled cluster method (RCCSDT) is calculated with a smaller basis set (aug-cc-pCVTZ) and added to the full interaction potential.
The importance of the scalar relativistic effects is assessed by performing coupled cluster calculations with the third-order Douglas-Kroll-Hess Hamiltonian~\cite{ReiherTCA06b}.
The uncertainty of the obtained molecular ion potential is evaluated by a systematic convergence analysis of the results obtained with different basis sets and methods.
The electronic structure calculations are performed with the \mbox{\textsc{molpro}} package of \emph{ab initio} programs~\cite{molprob}.

The resulting $\rtI{Li}$\,-\,\gt{Li}($2s_{1/2}$)~$X$\,$^2\Sigma$ ground state molecular ion potential $\tilde{V}_\IrAgM$ is displayed in Fig.~\ref{fig:InitialState}\fsl{a}.
For internuclear distances $R$ being much larger than the equilibrium distance $R_e$ of the molecular ion potential, $\tilde{V}_\IrAgM$ transitions into the polarization potential $\tilde{V}_\IrAg^{\gg}=-C_4/(2R^4)$.

\subsection{Ion-atom scattering calculations}
\label{subsec:SupplementScatteringCalc}
The scattering of the initial ion-atom wave packet $\Psi_\IrAgS(\bm{R},t=0)$ is described by the time-dependent Schr\"{o}\-dinger equation~\ref{eq:SEIrAgS} in which the nuclear Hamiltonian of the molecular ion is given by
\begin{equation}
\hat{H}_\IrAgM(\bm{R})=\frac{\IrAgPo^2}{2\mu_\IrAgM}+\tilde{V}_\IrAgM(R),
\label{eq:nHIrAgM}
\end{equation}
where $\IrAgPo$ is the momentum and $\mu_\IrAgM$ the reduced mass of the molecular ion system \IrAgM\ (see Fig.~\ref{fig:IrAgM}), and $\tilde{V}_\IrAgM(R)$ is the ground state molecular ion potential.

For the spatial and temporal propagation of the initial ion-atom wave packet, the radial part of the scattering Hamiltonian $\hat{H}_\IrAgM$ is represented on a Fourier grid with an adaptive step size~\cite{KokooulineJCP99b} and the angular part is expanded in terms of Legendre polynomials~\cite{TomzaMP13b}.
Without an electric field, the scattering Hamiltonian is diagonalized and its eigenstates and eigenenergies are used in propagation as given by Eq.~\ref{eq:PsiIrAgS}.
In an electric field, we propagate the wave packet using the Chebyshev propagator~\cite{KosloffARPC94b,TomzaPRA12b}.
To enable propagation times over microseconds and internuclear distances over \SI{e5}{\au}, we use a reduced ion-atom interaction potential which supports only several bound vibrational states and has the same long-range part and scattering length as the original potential.
We have found that reduced potentials supporting four or more vibrational states give the same scattering results.
The reduced interaction potential is obtained by adding a repulsive $C_{12}/R^{12}$ barrier to the original potential.
By adjusting $C_{12}$, we set the number of vibrational states and we control the scattering length.
We employ up to \num{11} partial waves and as many as \num{8192} radial grid points with a maximum internuclear distance of up to \SI{5e5}{\au}
The scattering calculations are performed with the developer version of the \textsc{qdyn} program package~\cite{qdynb}.

Without an electric field present, only $S$-wave scattering occurs and hence the scattered ion-atom wave packet is spherically symmetric [see inset in Fig.~\ref{fig:ScatteredWavepacket}\fsl{a} and \ref{fig:ScatteredWavepacket}\fsl{b}, respectively].
Electric stray fields $\mathcal{E}_\textrm{stray}$ during the scattering process admix higher partial waves to the scattered wave packet and thus break its spherical symmetry.
We performed scattering calculations showing that for any $\mathcal{E}_\textrm{stray}\leq\SI{0.1}{mV/cm}$ the introduced asymmetry in the scattered wave packet is small enough to still yield the same expansion velocity $\zeta$ and bound fraction $b$, and hence the same scattering length $\mathcal{A}$, as in the electric-field-free case [see Fig.~\ref{fig:ScatteredWavepacket}\fsl{c} and \ref{fig:ScatteredWavepacket}\fsl{d}, respectively].
An $\mathcal{E}_\textrm{stray}\leq\SI{0.1}{mV/cm}$ is experimentally achievable with suitable compensation schemes~\cite{Osterwalder99b,Huber14b}.

\parbox[c][0.6cm][c]{\columnwidth}{}

\makeatletter
\interlinepenalty=10000
\putbib[ias-ars]
\makeatother
\end{bibunit}

\end{document}